\documentclass[11pt,twoside]{article}
\usepackage{asp2010}

\resetcounters

\begin{document}

\title{ADS Labs - Supporting Information Discovery in Science Education}
\author{Edwin A. Henneken
\affil{Smithsonian Astrophysical Observatory, 60 Garden Street, MA 02138 Cambridge}}

\begin{abstract}
The SAO/NASA Astrophysics Data System (ADS) is an open access digital 
library portal for researchers in astronomy and physics, operated by the 
Smithsonian Astrophysical Observatory (SAO) under a NASA grant, 
successfully serving the professional science community for two decades. 
Currently there are about 55,000 frequent users (100+ queries per year), 
and up to 10 million infrequent users per year. Access by the general public 
now accounts for about half of all ADS use, demonstrating the vast reach 
of the content in our databases. The visibility and use of content in the 
ADS can be measured by the fact that there are over 17,000 links from 
Wikipedia pages to ADS content, a figure comparable to the number of 
links that Wikipedia has to OCLC’s WorldCat catalog. The ADS, through its 
holdings and innovative techniques available in ADS Labs (http://adslabs.org), 
offers an environment for information discovery that is unlike any 
other service currently available to the astrophysics community. Literature 
discovery and review are important components of science education, 
aiding the process of preparing for a class, project, or presentation. The 
ADS has been recognized as a rich source of information for the science 
education community in astronomy, thanks to its collaborations within 
the astronomy community, publishers and projects like ComPADRE. One 
element that makes the ADS uniquely relevant for the science education 
community is the availability of powerful tools to explore aspects of the 
astronomy literature as well as the relationship between topics, people, 
observations and scientific papers. The other element is the extensive 
repository of scanned literature, a significant fraction of which consists 
of historical literature.
\end{abstract}

\section{Introduction}
Information discovery is central to many activities in life, from finding restaurants while attending a conference to making strategic decisions in a big company. In science, information discovery is, for example, used to stay up-to-date, doing literature research, and it is crucial in the process of scientific research itself. In a world where online information is generated 24 hours a day, 7 days a week, this journey of discovery can easily become a daunting task. We need powerful discovery tools to help us on this journey. General search engines support a low level of information retrieval, sufficient to get a general idea, but when you are looking for technical material in a rich metadata environment, you need specialized digital libraries. The SAO/NASA Astrophysics Data System (ADS) is such a digital library (\citet{kurtz00}, \citet{henneken11a}). It has very successfully served the astronomy and physics community for almost 20 years, free of charge. In order to support a richer and more efficient information discovery experience, we created the ADS Labs environment, in which we expose our users to new search paradigms and tools, to better support our community's research needs. In the following we will argue that the ADS Labs environment will prove to be a useful tool for people involved in science education and outreach.

\section{Using ADS Labs}
General search engines return so many results that it quickly consumes excessive amounts of time to parse all these results and determine if any of these are relevant in a science education environment. As a result, instructors often return to resources they have been using time and time again, while missing out on a wealth of new material continuously being added through a broad spectrum of resources. A specialized digital library is a highly efficient tool to locate these resources. Through its contents and functionality, ADS Labs will prove to be such tool.

In addition to publications relevant for scientific research, the ADS repository also contains a set of journals that are directly relevant to people involved in science education. Examples of such journals are: {\it Astronomy Education Review}, {\it American Journal of Physics}; {\it The Science Teacher}; {\it Journal of Science Education and Technology}; {\it Journal of Science Teacher Education}; {\it International Journal of Science Education}; {\it Research in Science Education}; {\it Science \& Education}; {\it Spark, the AAS Education Newsletter}. A large portion of the astronomical research of the 19th and early 20th centuries was reported in publications written and published by individual observatories. Many of these collections were not widely distributed and complete sets of these volumes are now, at best, difficult to locate. This material is often requested by amateur astronomers and researchers, because the observatory report is the only published record of the research and observations. This makes these publications a great resource for classes that have a component dealing with the history of astronomy. We have a large number of historical publications (mostly from microfilm) in our repository of scanned literature (\citet{thompson07}).

\begin{figure}[!ht]
  \plotone{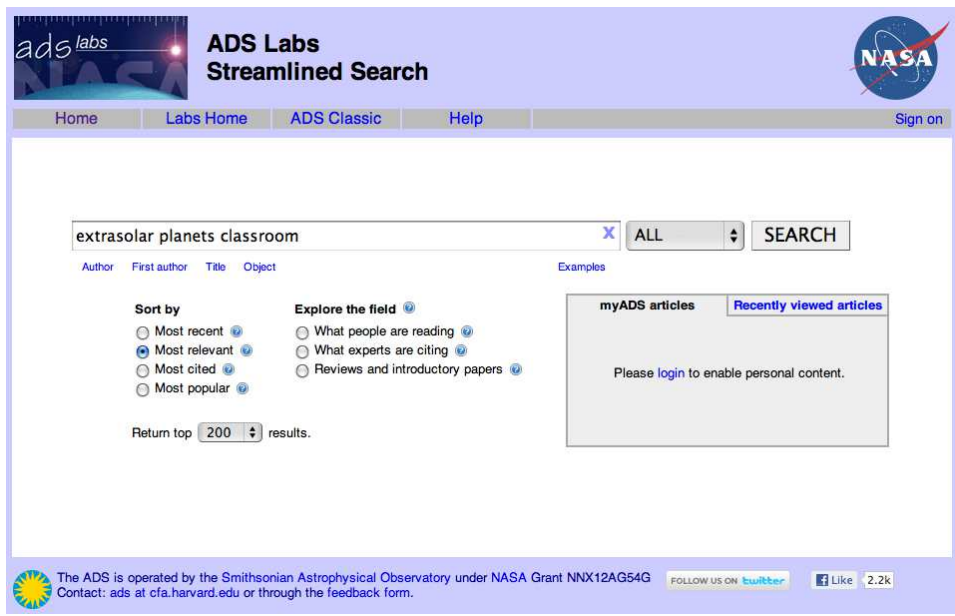}
  \caption{Streamlined Search. This query example will find publication that have the terms {\it extrasolar planets classroom} in their title, sorted by ``relevancy''. The ``AST'' next to the search button refers to the fact that the entire database is searched (not just astronomy, for example).}
\label{screenshot:streamlined}
\end{figure}

The functionality that makes ADS Labs such a powerful tool is a combination of being able to specify ahead of time what kind of results you are interested in, and the ability to efficiently filter the results afterwards, using facets. We will illustrate this using an example. Imagine you are looking for publications describing or on discussing extrasolar planets in a classroom environment. Figure~\ref{screenshot:streamlined} shows how you could start this search, using the ``streamlined search'' of ADS Labs. 
In the search box we specify ``extrasolar planets classroom'', as search scope we select ``ALL'' (next to the search button), to search the entire ADS repository, and we specify ``most relevant'' for sorting. This sorts on a combination of several indicators, including date, position of the query words in the document, position of the author in the author list, citation statistics and usage statistics (this is how many popular search engines rank).
\begin{figure}[!ht]
  \plotone{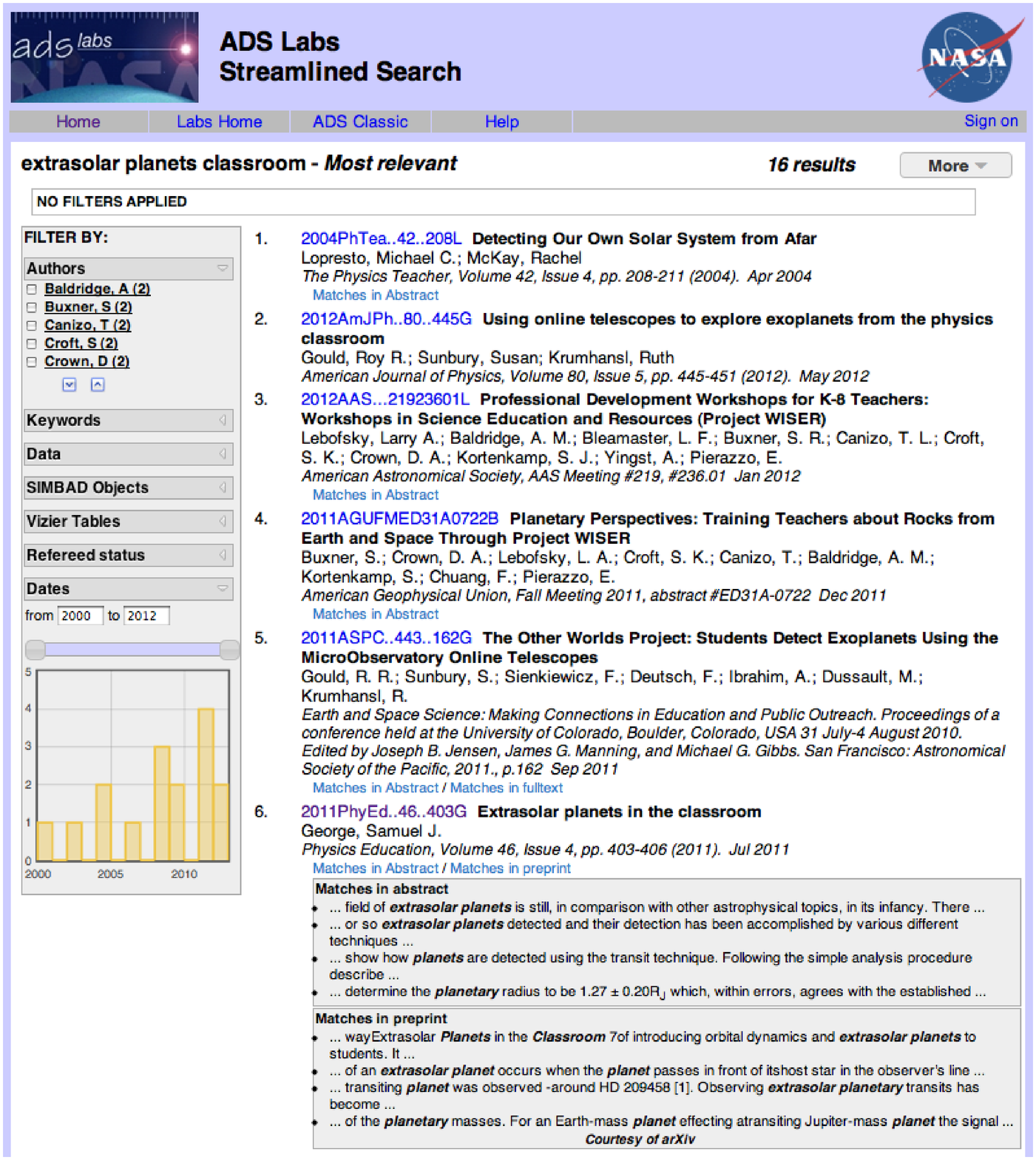}
  \caption{Results page. The results page consists of a list of publications with a panel of facets on the left. Via snippets the user is given a view inside the abstract or full text to see matches of the query terms.}
\label{screenshot:results}
\end{figure}
This example query generates the results list shown in figure~\ref{screenshot:results}. The results page consists of a list of publications with a panel of facets on the left, which offer an efficient way of further filtering the results. Besides serving as a filter, the author facet summarizes the people active in the field defined by the query. The author facet also shows all the spelling variants by which author names occur in the ADS repositories (by clicking on the author name), which can assist in filtering out different authors with the same initials (where e.g. the first name is spelled out). The diagram below the facets shows the number of publications as a function of year, providing a measure for the activity in a field. In this way the facets, besides serving as filters, also provide you with valuable information. The ``Data'' facet provides you with an overview of available data products, if available. Every entry in the results list has a potential option to look inside the publication. If an abstract is available, the ``Matches in Abstract'' link will open a snippet showing the matches in the abstract of the query terms. The link ``Matches in fulltext'' will show these matches in the full text version of the publication (which could be the preprint version from arXiv). The results page also provides the menu ``More'' (in the upper right corner) that contains tools to further explore the results. The ``author network'' allows one to visualize collaborations between authors and further filter the search results by selecting within the nodes in the network. The paper network allows one to visualize the relationships between papers and further filter the search results by selecting within the nodes in the network. If the publications in the results list have astronomical objects identified in them, you can visualize these using the ``Sky Map'' option. This uses Google Sky. The sky map can also be used for further filtering. The ``Word Cloud'' visualizes the most relevant terms found in the list of results and further filter the list by selecting interesting terms from the cloud. The word cloud is only indirectly based on word frequencies: the word frequencies in the astronomy corpus are subtracted, so that relatively frequent words stand out. The ``Metrics'' page provides an overview of bibliometric indicators, based on the publications in the results list. This overview can be exported in Excel format. 
\begin{figure}[!ht]
  \plotone{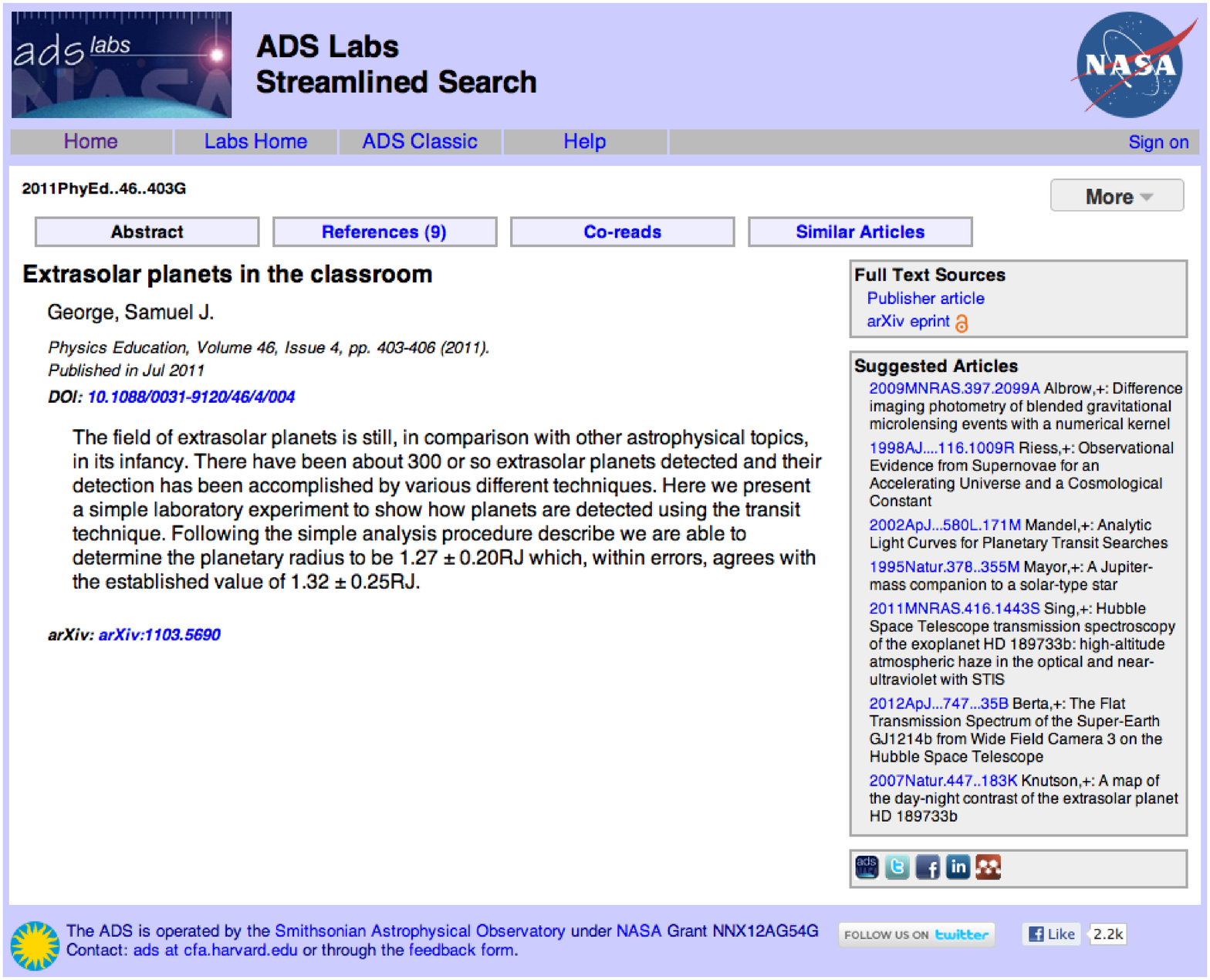}
  \caption{Abstract page. This page provides the user with a concentrated description of an article.}
\label{screenshot:abstract}
\end{figure}
The next level is the abstract page (see figure~\ref{screenshot:abstract}). The principal function of this page is to provide the user with a concentrated description of an article, sufficient for the user to decide whether or not to download and read it. This page provides a view on the basic metadata (authors, affiliations, title, abstract). It also provides links to the full text (including open access versions) and opportunities to share the abstract on various social media sites, or save it in an ADS Private Library (available for users with an ADS account, see below). When sufficient data is available, a set of ``Suggested Articles'' is provided, which is generated by a recommender system (\citet{henneken11b}). Depending on availability, the abstract view provides access to the bibliography of the publication (those papers cited by the publication, for which there is a record in the ADS), an overview of publications citing the publication, a list of papers most frequently read in conjunction with this publication (``Co-reads'') and a list of papers that are similar to this publication, based on word similarity (``Similar Articles''). 

Figure~\ref{screenshot:streamlined} has a area on the right with two tabs (``myADS articles'' and ``Recently viewed articles''). These relate to the existence of user accounts in the ADS. The ADS offers the option to create a login, providing its users with the possibility to personalize the service. This includes access to a powerful alert service called ``myADS'' (see~\citet{henneken07}), the creation of ``Private Libraries'' (which are essentially baskets in which users can store, and annotate, links to publications) and a way to specify a ``Library Link Server'', allowing access to full text using institutional subscriptions. This makes the ADS portable, because it provides you with online access to the full text of articles from anywhere in the world. When a user is logged in to their account, visiting the streamlined search will result in displaying the most recent content for their ``daily myADS'' service and an overview of their most recently viewed records.

As an illustration of finding historical material, consider the following example: you are interested in the history and use of an instrument called the ``mural circle''. When you run the query ``\"mural circle\" 1800-1850'' in the streamlined search, you will find about two dozen results. These results show that in this period there were such instruments at the Madras Observatory, Royal Greenwich Observatory, Armagh Observatory, U.S. Naval Observatory and Cape Observatory. Most publications in this results list are available in PDF format. 

\section{Concluding Remarks}
Through the publications in its holdings and the user-friendly, intuitive streamlined search, the ADS is a useful instrument in the tool box of search engines for professionals involved in science education and outreach. We do realize that this is a group of users with requirements and needs that, in some aspects, differ significantly from those that have traditionally been using the ADS. We would love to get feedback and suggestions to help us optimize the search experience for all our users. Our users are a big part of our curation efforts, so if you encounter material in our database relevant for science education, but not flagged as such, we would love to hear from you as well. Feedback should be sent to ads@cfa.harvard.edu.

\acknowledgements The ADS is funded by NASA grant NNX12AG54G.

\end{document}